# How Good is Artificial Intelligence at Automatically Answering Consumer Questions Related to Alzheimer's Disease?

Krishna B. Soundararajan[1], Sunyang Fu, MHI[2], Luke A. Carlson[2], Rebecca A. Smith[2], David S. Knopman, MD[2], Hongfang Liu, PhD[2], Yanshan Wang, PhD[2]
[1]Carnegie Mellon University, Pittsburgh, PA, USA; [2]Mayo Clinic, Rochester, MN, USA

**Background and Motivation**

Alzheimer's Disease (AD) is the most common type of dementia, comprising 60-80% of cases[1]. There were an estimated 5.8 million Americans living with Alzheimer's dementia in 2019, and this number will almost double every 20 years[1]. The total lifetime cost of care for someone with dementia is estimated to be $350,174 in 2018, 70% of which is associated with family-provided care[1]. Most family caregivers face emotional, financial and physical difficulties. As a medium to relieve this burden, online communities in social media websites such as Twitter, Reddit, and Yahoo! Answers provide potential venues for caregivers to search relevant questions and answers, or post questions and seek answers from other members. However, there are often a limited number of relevant questions and responses to search from, and posted questions are rarely answered immediately. Due to recent advancement in Artificial Intelligence (AI), particularly Natural Language Processing (NLP), we propose to utilize AI to automatically generate answers to AD-related consumer questions posted by caregivers and evaluate how good AI is at answering those questions. To the best of our knowledge, it is rare in the literature applying and evaluating AI models designed to automatically answer consumer questions related to AD.

**Methods**

There are several AI models that have recently shown promising results for NLP tasks. Most of them are language models based on the Transformer architecture, among which BERT[2] and GPT-2[3] have achieved state-of-the-art performance for most NLP tasks. Since BERT generates sentences that are more diverse but of slightly worse quality than GPT-2[4], we utilized GPT-2 (medium model with 345M parameters) to automatically generate responses to consumer questions. GPT-2 is an unsupervised deep learning model trained from 40 GB of Web texts, which generates coherent and semantically relevant textual responses for prompted texts that can be served as inferred answers. Since the questions are related to AD and care of disease, we additionally trained a transfer learning based GPT-2 model, named EduGPT-2, on a dataset of patient education materials from Mayo Clinic which contains more than 9k patient education documents related to diseases, symptoms, causes, treatment, preparation for and care after procedures, etc. To create a consumer question corpus, we crawled 1000 question titles under the topic "Alzheimer's disease" from Yahoo! Answers. Since many questions are fragmented and poorly formed, we asked two annotators with medical background to manually classify these questions into two categories "well-formed" and "poorly-formed". Then we randomly selected 30% questions from the "well-formed" category, and applied GPT-2 and EduGPT-2 to generate responses separately. We asked two annotators to evaluate the responses by giving a relevance score between 1 and 4 with 4 being the most relevant response and 1 being completely irrelevant to the question. In order to avoid bias, the responses generated for each question by GPT-2 and EduGPT-2 were shuffled so that the annotators would not know which response corresponded to which model.

**Results**

Based on the question categorization, there is a total of 277 "well-formed" questions. 84 questions (30% of 277) were randomly selected for the evaluation of automatically generated answers. Figure 1 depicts the annotation results for the responses to 84 questions generated by two AI models. Most responses are relevant to the questions and have relevance scores of 4 or 3. We utilized the average relevance score of the two annotators for each question to evaluate the AI models. The mean relevance score of responses generated by GPT-2 and EduGPT-2 are 3.0 and 2.8, with inner annotator agreement scores of 0.43 and 0.22, respectively.

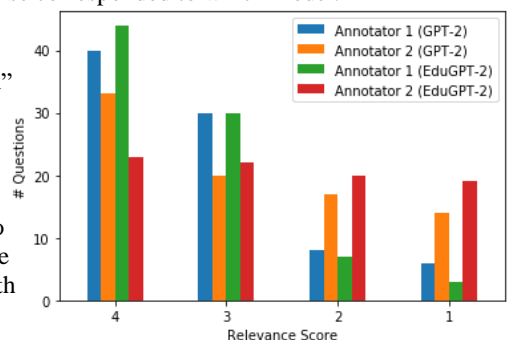

**Figure 1.** Annotation results.

**Conclusion**

Annotation results showed the feasibility of applying AI models to automatically answer consumer questions related to AD. The results are also interesting in that the original GPT-2 performs slightly better than EduGPT-2 based on transfer learning, which may be due to the slightly poor generalizability of EduGPT-2 and needs further study.